\documentclass[11pt, showpacs,showkeys]{revtex4} 
\usepackage{hyperref}
\usepackage{amsmath}
\usepackage{amsfonts,epsfig}
\usepackage{amssymb}

\def\dps{\displaystyle}

\def\no{\noindent}
\def\non{\nonumber\\}
\def\be{\begin{equation}}
\def\ee{\end{equation}}
\def\bea{\begin{eqnarray}}
\def\eea{\end{eqnarray}}
\def\bear{\begin{eqnarray}}
\def\ear{\end{eqnarray}}
\def\bei{\begin{itemize}}
\def\eei{\end{itemize}}
\def\bee{\begin{enumerate}}
\def\eee{\end{enumerate}}

\def\Eins{{\mathchoice {\rm 1\mskip-4mu l} {\rm 1\mskip-4mu l}
{\rm 1\mskip-4.5mu l} {\rm 1\mskip-5mu l}}}
\def\Z{{\mathchoice {\hbox{$\sf\textstyle Z\kern-0.4em Z$}}
{\hbox{$\sf\textstyle Z\kern-0.4em Z$}}
{\hbox{$\sf\scriptstyle Z\kern-0.3em Z$}}
{\hbox{$\sf\scriptscriptstyle Z\kern-0.2em Z$}}}}
\def\e{\,{\rm e}}

\begin{document}

\title{Full mass range analysis of the QED effective action for an $O(2)\times O(3)$ symmetric field}

\pacs{11.10.-z, 11.15.-q, 12.20.-m, 12.20.Ds}
\keywords{QED, effective action, $O(2)\times O(3)$ symmetric fields, partial-wave-cutoff method.}

\author{Naser Ahmadiniaz$^{a}$, Adolfo Huet$^{a,b}$, Alfredo Raya$^a$ and Christian Schubert$^{a}$}

\affiliation{$^a$ \it 
Instituto de F\'{\i}sica y Matem\'aticas
\\
Universidad Michoacana de San Nicol\'as de Hidalgo\\
Edificio C-3, Ciudad Universitaria\\
C.P. 58040, Morelia, Michoac\'an, M\'exico\\}
\affiliation{$^b$\it Facultad de Ciencias F\'{\i}sico-Matem\'aticas, \\
Universidad Michoacana de San Nicol\'as de Hidalgo\\
 Avenida Francisco J. M\'ujica S/N, C.P. 58060, Morelia,
Michoac\'an, M\'exico\\
$\phantom{x}$\\
{naser/raya/schubert@ifm.umich.mx, adolfo.huet@gmail.com}}%\affiliation{$^c$
%{\it 
%Institutes of Physics and Mathematics, Humboldt-Universit\"at zu Berlin,\\
%Unter den Linden 6, 10099 Berlin, Germany
%}}

%{\it
%Dipartimento di Fisica, Universit\`a di Bologna and INFN, Sezione di Bologna,\\
%Via Irnerio 46, I-40126 Bologna, Italy}}

%$\emailAdd{naser@ifm.umich.mx}
%\emailAdd{adolfo.huet@gmail.com}
%\emailAdd{raya@ifm.umich.mx}
%\emailAdd{schubert@ifm.umich.mx}
%\date{\today}

\begin{abstract}
An interesting class of background field configurations in quantum electrodynamics (QED) are the $O(2)\times O(3)$ symmetric fields, originally introduced by S.L. Adler in 1972.
Those backgrounds have some instanton-like properties and yield a one-loop effective action that is highly nontrivial, but amenable to numerical calculation. 
Here, we use the recently developed ``partial-wave-cutoff method'' for a full mass range numerical analysis of the effective action for the ``standard'' $O(2)\times O(3)$ 
symmetric field, modified by a radial suppression factor. 
At large mass, we are able to match the asymptotics of the physically renormalized effective action against the leading two mass levels of the inverse mass expansion. 
For small masses, with a suitable choice of the renormalization scheme we obtain stable numerical results even in the massless limit. 
We analyze the $N$ - point functions in this background and show that, even in the absence of the radial suppression factor, the two-point contribution to the
effective action is the only obstacle to taking its massless limit.  
The standard $O(2)\times O(3)$ background leads to a chiral anomaly term in the effective action, and 
both our perturbative and nonperturbative results strongly suggest that the small-mass asymptotic behavior of the effective action 
is, after the subtraction of the two-point contribution, dominated by this anomaly term as the only source of a logarithmic mass dependence. 
This confirms a conjecture by M. Fry. 
\end{abstract}
%\begin{document}
\maketitle
%\flushbottom

\section{Introduction}
\renewcommand{\theequation}{1.\arabic{equation}}
\setcounter{equation}{0}

The one-loop effective action in QED can be calculated analytically only for a limited variety of background fields, 
such as the constant field-strength case \cite{Heisenberg:1935qt,Schwinger:1951nm}, and some special inhomogeneous configurations \cite{Dunne:2004nc}.
For generic backgrounds, numerical or other approximation methods have to be used. In particular, taking the loop scalar or electron
mass to be either zero or large generally leads to simplifications. This is particularly so for the latter case since, at large mass, the
effective action can be reduced to its heat kernel expansion, which is simple in structure and
for whose computation powerful methods exist  \cite{novikov,ball,lepash,5}.
However, it is fair to say that, even at the numerical level, presently
there is still no method available that would allow one to obtain reliable 
results for the effective action for arbitrary masses and in a generic background 
(the ``worldline Monte Carlo'' approach
\cite{gielan,gisava} may ultimately provide such a formalism, although it seems too early to tell). 
For radially separable backgrounds, on the other hand, during the last few years
the so-called ``partial-wave-cutoff method'' \cite{Dunne:2004sx, Dunne:2005te, Dunne:2006ac,Dunne:2007mt,Hur:2008yg} has been
developed, which seems to have all the properties one might request of such a numerical method.
This class of backgrounds, although still far from generic, includes, e.g., instantons, monopoles and vortices. 
The method, originally invented for the case of the quark determinant in an instanton background,
is based on a decomposition of the relevant one-loop operator into partial waves of definite angular momentum, and a separation into
low and high angular momentum contributions, where the former are computed using the Gel'fand-Yaglom method, and
the latter in a WKB expansion. It principally applies to the scalar loop, but can be extended to the spinor loop case
for certain backgrounds. An example of this is the instanton where, by self-duality, the spinor effective action can be reduced to the scalar one \cite{'tHooft:1976fv,Jackiw:1977pu,Brown:1977bj,Carlitz:1978xu}.

In \cite{Dunne:2011fp} G.V. Dunne et al. initiated the application of this method to the important class of $O(2)\times O(3)$ symmetric fields, first introduced
by S.L. Adler in \cite{Adler:1972qq, Adler:1974nd} and studied later by a number of authors 
\cite{itpazu,bizp,Bogomolny:1981qv, Bogomolny:1982ea}. These backgrounds can be defined (in  Euclidean metric) as 

\begin{equation}
A_{\mu} (x) = \eta^3_{\mu \nu} x_\nu g(r) \; ,
\label{deffield}
\end{equation}
where $ \eta^3_{\mu \nu}$ is a 't Hooft symbol \cite{'tHooft:1976fv},
and $g(r)$ a radial profile function. 
Such backgrounds provide a good testing ground since, on one hand,
they still permit a reduction from the spinor to the scalar loop case, 
while on the other hand, the profile function $g(r)$ is, to a good extent,  arbitrary,  thus leading to a large class of models.
In \cite{Dunne:2011fp}, the ``partial-wave-cutoff method'' was applied to various profile functions, in the full mass range,
and the results compared to the large mass expansion,
as well as to the derivative expansion.
In all cases, the method was found to be in good agreement with the large mass approximation, and in
some cases, also the derivative expansion could be used to check agreement in the small mass regime.
In the present paper, we continue the investigation started in \cite{Dunne:2011fp} in two directions.

First, in \cite{Dunne:2011fp}, the renormalization of the effective action had been done using an unphysical renormalization condition, designed to yield a finite zero-mass limit. The asymptotic behavior for large mass then is dominated by an unphysical logarithm that made it difficult to numerically test this behavior beyond that logarithmic term. Here, we instead consider the
physically renormalized effective action, which has a logarithmic divergence at small, but not at large mass. This enables us
to probe the large mass behavior in deeper detail than was achieved in \cite{Dunne:2011fp}. 
Specifically, we are able to numerically verify the two leading
mass levels in the large mass expansion of the physical effective action.
Since not all of the relevant terms in this expansion seem to be available in the literature, 
and moreover the coefficients depend on the chosen operator basis, 
we also present here their calculation from scratch, using the
worldline path integral formalism along the lines of \cite{5}. 

Second, this class of $O(2)\times O(3)$ symmetric backgrounds has been extensively studied by M. Fry 
\cite{fry2003,fry2006,fry2010} in a long-term effort to demonstrate that, as in the case of 1+1 dimensional QED \cite{fry2000,fryPRD62}, also in the
four-dimensional case the
small $m$ behavior of the effective action is, after subtraction of the two- and four-point contributions, 
dominated by a $\ln m$ coming from the chiral anomaly term $\sim \int d^4x F_{\mu\nu}\tilde F^{\mu\nu}$, whenever such a term is present.
A simple test case for this conjecture in the class of backgrounds defined by (\ref{deffield}) would be the profile function

\bear
g(r) = \frac{\nu}{r^2+\rho^2}
\label{gstandard}
\ear
where $\nu,\rho$ are positive constants. The background (\ref{deffield}) with this profile function 
will be called the ``standard $O(2)\times O(3)$ symmetric background'' in the following. 
Our numerical method does not really allow us
to treat this case as it stands, since it has insufficient radial fall-off. 
Even though, we will provide strong support for this hypothesis by supplying the profile function
(\ref{gstandard}) with a radial suppression factor $\e^{-\alpha r^2}$, and studying the double limit of small $m$ and small $\alpha$. 
Combining a perturbative and nonperturbative approach, we 
will show that the appropriately renormalized effective action remains finite in the small $m$ limit for any positive
value of $\alpha$, and that the only obstacle preventing one to take the double limit $m,\alpha\to 0$ resides in the perturbative
two-point contribution to the effective action. 

The paper is organized as follows. In Sect.~\ref{The background}, we review the properties of the
$O(2)\times O(3)$ background, and the predictions made in \cite{fry2006,fry2010} about its effective action.
Section~\ref{method} presents a review of the partial-wave-cutoff method.
In Section \ref{perturbative} we study the perturbative $N$ - point functions in the standard $O(2)\times O(3)$ symmetric background
modified by the radial suppression factor. This Section contains also the calculation of the leading and subleading terms in
the inverse mass expansion. 
In Sect.~\ref{nonperturbative}, we present our numerical results for the effective action. 
Our conclusions are presented in Sect.~\ref{conc}.

\section{The $O(2)\times O(3)$ field and its effective action}
 \label{The background}
\renewcommand{\theequation}{2.\arabic{equation}}
\setcounter{equation}{0}

Let us start with some general facts on the effective action in four-dimensional spinor QED (we work in the Euclidean space
throughout). This effective action can be written either in terms of the Dirac operator or
its square as

\begin{eqnarray}
\Gamma[A] = -\ln \, \det\left( \displaystyle{\not} D+m \right)
=-\frac{1}{2}\ln\,\det\left(-\displaystyle{\not} D^2 +m^2\right)\;,
\label{oneloop}
\end{eqnarray}
Here, $\displaystyle{\not} D
= \gamma_\mu( \partial_\mu + i eA_\mu (x) )
$ is the Dirac operator in 4-dimensional spacetime, and $A_\mu(x)$ is the classical background gauge field.
We set $e=1$, unless explicitly indicated.
%To avoid undue repetition, we discuss only the spinor case  in this section; the simpler scalar case 
%has been treated in detail in \cite{Dunne:2007mt}. 
We use a standard representation of the  Dirac matrices as in \cite{Jackiw:1977pu}, and obtain the following
chiral form for the squared Dirac operator:
\begin{eqnarray}
-\displaystyle{\not} D^2 + m^2 =
\begin{pmatrix}
m^2-D_\mu^2+\frac{1}{2}F_{\mu\nu} \bar{\eta}^a_{\mu\nu}\,\sigma_a & 0  \cr
0 & m^2-D_\mu^2+\frac{1}{2} F_{\mu\nu} \eta^a_{\mu\nu}\,\sigma_a
\end{pmatrix}\;.
\end{eqnarray}
%where the $\eta^a_{\mu \nu},\bar\eta^a_{\mu \nu}$'s are 't Hooft symbols.
Thus, we have a chiral decomposition of the effective action,

\bear
\Gamma[A] &=& -\frac{1}{2}\ln\,\det\left(-D^2 +m^2+\frac{1}{2}F_{\mu\nu} \bar{\eta}^a_{\mu\nu}\,\sigma_a \right)
-\frac{1}{2}\ln\,\det\left(-D^2 +m^2+\frac{1}{2}F_{\mu\nu} {\eta}^a_{\mu\nu}\,\sigma_a \right)
\non
&=:& \Gamma^{(+)}[A] + \Gamma^{(-)}[A] \;.\non
\label{decompchiral}
\ear
We can also write

\begin{eqnarray}
\Gamma[A] &=& 2\Gamma^{(\pm)}[A]
\mp\left(\Gamma^{(+)}[A]-\Gamma^{(-)}[A]\right)\;.
\label{choice}
\end{eqnarray}
Further, it is known that after renormalization,
the difference of the renormalized effective action for the two chiralities is related to the chiral anomaly as

\begin{eqnarray}
\Delta\Gamma_{\rm ren}[A] &\equiv& \left(\Gamma^{(+)}_{\rm ren}[A]-\Gamma^{(-)}_{\rm ren}[A]\right)
=\frac{1}{2}\frac{1}{(4\pi)^2}\ln \left(\frac{m^2}{\mu^2}\right)\int d^4 x\, F_{\mu\nu} \tilde F_{\mu\nu}\;.
\label{difference}
\end{eqnarray}
Therefore, for the computation of the full spinor effective action, it is sufficient to evaluate either 
$\Gamma_{\rm ren}^{(+)}[A]$ or $\Gamma_{\rm ren}^{(-)}[A]$
and the chiral anomaly term \cite{Hur:2010bd}. This fact is computationally quite relevant,
since the contribution of one chirality might be significantly easier to compute than the 
other one. 

\no
We consider a field of the form (\ref{deffield}),
%\begin{equation}
%A_{\mu} (x) = \eta^3_{\mu \nu} x_\nu g(r) \; ,
%\label{deffield2}
%\end{equation}
 where $g(r)$ is a radial profile function. For this background, the negative chirality part of the Dirac operator takes a simple form,
 \begin{eqnarray}
m^2-D_\mu^2+\frac{1}{2} F_{\mu\nu} \eta^a_{\mu\nu}\,\sigma_a = m^2-D_\mu^2+\left(4g(r)+r\,g'(r)\right)\sigma_3 \;.
\label{negative}
\end{eqnarray}
Hence, we use (\ref{choice}) in the form

\begin{eqnarray}
\Gamma_{\rm ren}[A] &=& 2\Gamma^{(-)}_{\rm ren}[A] + \Delta\Gamma_{\rm ren}[A]\;.
\label{choose}
\end{eqnarray}
The field strength tensor for the background (\ref{deffield}) is

\begin{equation}
F_{\mu \nu}(x) = - 2 \eta_{\mu \nu}^3 g(r) - \frac{g'(r)}{r} \left(\eta_{\mu \sigma}^3 x_\nu x_\sigma   -  \eta_{\nu \sigma}^3 x_\mu x_\sigma\right)\;.
\label{f}
\end{equation}
From (\ref{difference}), we see how the choice of $g(r)$  determines the presence or absence of zero-modes.
To be  specific, the number of zero-modes is counted by 

\begin{eqnarray}
\frac{1}{2} \int_0^\infty dr \; r^3 (F_{\mu \nu}\tilde F_{\mu \nu})  &=&
\int_0^\infty dr \; r^3 (8 g(r)^2 + 4 g(r)g'(r) r ) 
%&=& \int_0^\infty dr \; \frac{d}{dr} (2 g(r)^2 r^4 )  \, , \nonumber \\
= 2 (g(r) r^2)^2 \Bigg|_0^{\infty}  \, .
\end{eqnarray}
Therefore, as long as $g(r)$ falls faster than $1/r^2$, there are no zero-modes. 
As was already mentioned in the introduction, the partial-wave-cutoff method --which we wish to
use-- is not guaranteed to work well in  the standard case $g(r) = 1/(r^2+\rho^2)$ due to insufficient radial fall-off. 
This suggests to study the following more general family of profile functions~\cite{Dunne:2011fp}:

\begin{equation} \label{g2}
g (r) \equiv \nu\,  \frac{e^{-\alpha r^2}}{\rho^2 + r^2} \, , 
\end{equation}
where  $\nu$,
$\rho$ and $\alpha$ are parameters that control the amplitude, steepness, and range of the potential.  
For this profile
function, the choice of $\alpha$ produces one of  the following  cases:

\begin{eqnarray}
\alpha > 0   &\Longrightarrow& \quad \int d^4 x F_{\mu \nu} F_{\mu \nu} < \infty
\quad , \quad \int d^4 x F_{\mu \nu} \tilde F_{\mu \nu} = 0 \;,\label{alphapos}\\
\alpha = 0   &\Longrightarrow& \quad \int d^4 x F_{\mu \nu} F_{\mu \nu} \to \infty 
\quad , \quad 0 < \Big| \int d^4 x F_{\mu \nu} \tilde F_{\mu \nu} \Big| < \infty  \;.
\label{alphazero} 
\end{eqnarray} 
Thus, we have zero modes only for $\alpha = 0$, corresponding to the standard case. 
It will also be useful to note that then

\begin{eqnarray}
\frac{1}{2} \int_0^\infty dr \; r^3 F_{\mu \nu} F_{\mu \nu}  &=&
\int_0^\infty dr \; r^3 (8 g(r)^2 + 4 g(r)g'(r) r +
g'(r)^2 r^2 ) \, ,\nonumber  \\
&=&  \int_0^\infty dr \; r^3 F_{\mu \nu} \tilde F_{\mu \nu} + \int_0^\infty dr \; r^3 ( g'(r)^2 r^2 ) \;,
%\, , \nonumber \\
%&=& {\rm {finite}} \quad + \quad \sim \int_0^\infty  \; \frac{dr}{r} \, , \nonumber \\
%&=& \lim_{R \to \infty} \ln R \, .
\end{eqnarray} 
where the second integral diverges logarithmically.
In the present paper, we will set $\rho =1$ throughout,  and also $\nu =1$, unless explicitly stated otherwise.
$\alpha$ will be a small  positive number effectively serving as an IR cutoff.

Finally, let us summarize the conclusions reached at in~\cite{fry2006,fry2010} about the small mass limit of the spinor QED effective action
for the standard case $\alpha = 0$:

\begin{enumerate}

\item
Let $\cal R$ denote the (scheme independent) effective action obtained after subtraction of the two-point contribution.

\item
There is evidence that $\cal R$ behaves for small $m$ as

\bear
{\cal R} \, {\sim} \, \frac{\nu^2}{4} \ln m^2 + {\rm less\,}{\rm singular}\,{\rm in}\, m^2\;.
\label{limfry}
\ear 

\item
The logarithmic term in (\ref{limfry}) is determined entirely by the chiral anomaly, given for our background by

\bear
-\frac{1}{(4\pi)^2}  \int d^4 x F_{\mu \nu} \tilde F_{\mu \nu}  = \frac{\nu^2}{2}\;.
\label{anomalyint}
\ear

\end{enumerate}

\noindent
In this paper, we provide strong support for these statements.
Further, an important application of (\ref{limfry}) is the search for a nontrivial zero of the determinant 

\bear
{\rm ln} {\rm det}_5 := {\cal R} - \Pi_4\;,
\label{defdet5}
\ear
where $\Pi_4$ is the four-point contribution to $\cal R$ \cite{fry2003,fry2006,fry2010}. In this connection, it is useful to know whether
the four-point contribution itself adds to the logarithmic singularity of the massless limit,
$\Pi_4 \sim C \,{\rm ln}\, m$. As a corollary of our study of the $N$ - point functions in Section \ref{perturbative} 
we will settle this detail, that had been
left open in the analysis of \cite{fry2006}, by showing that $C=0$.

%%%%%%%%%%%%%%%%%%%%%%%%%%%%%%%%%%%%%%%%%%%%%%%%%%%%%%%%%%%%%%%%%%%%%%%%%%%%%%%%%%%%%%%%%%%%%%%%%%%%%%%%%%%%%%%%%%%%%%%

\section{The partial-wave-cutoff method}
\label{method}
\renewcommand{\theequation}{3.\arabic{equation}}
\setcounter{equation}{0}

We now explain the application of the partial-wave-cutoff method to the above family of backgrounds, following \cite{Dunne:2011fp}
closely. We concentrate on the negative chirality sector of the effective action. 

After decomposing the negative chirality part of the Dirac operator into partial-wave radial operators with 
quantum numbers $l$ and $l_3$, we can apply the partial-wave cutoff method. The effective action is given by  a sum over angular momentum
eigenmodes:

\begin{eqnarray}
2 \Gamma^{(-)} &=& -\sum_{s = \pm} \; \sum_{l =
0, \frac{1}{2} , 1, \ldots}^\infty \; \Omega(l) \sum_{l_3 = -l}^{l}
\ln \left( \frac{ {\rm{det}}(m^2+\mathcal{H}_{(l, l_3,
s)})}{{\rm{det}}(m^2+\mathcal{H}_{(l, l_3, s)}^{{\rm{free}}})} \right) \,, 
\label{angsum}
\end{eqnarray}
where $\Omega(l) =  (2 l +1) $
is  the degeneracy factor, and the
$s$ sum comes from adding the contributions of each spinor
component.
For high values of $l$ (high-modes), we use a WKB expansion of   $ \ln [ {\rm{det}}(m^2+\mathcal{H}_{(l, l_3,s)})]$. 
However, this expansion does not apply near $l=0$ (low-modes).
The basic idea of the partial-wave-cutoff-method is to separate
 the sum over the the quantum number $l$ into a low partial-wave contribution, each term of which is computed using the (numerical) Gel'fand-Yaglom method, and a high partial-wave contribution, whose sum is computed analytically using WKB. 
Then, we apply a regularization and renormalization procedure and combine these two contributions to yield the finite and renormalized effective 
action \cite{Dunne:2004sx,Dunne:2005te,Dunne:2006ac,Dunne:2007mt,Hur:2008yg}.

\noindent
In a basis of angular momentum eigenstates, from (\ref{negative}), we get the following partial-wave operators: 

\begin{eqnarray} \label{radial1} 
m^2+\mathcal{H}_{(l, l_3, s)} = -\Big[ \partial_r^2  + \frac{4 l + 3}{r} \partial_r  - r^2 g(r)^2 - 4 g(r) l_3   - m^2
\mp (4 g(r) + r g'(r) ) \Big] \;.
\label{pwo}
\end{eqnarray}
The quantum number $l$ takes half-integer values: $l=0, \frac{1}{2}, 1, \frac{3}{2}, \dots$, while $l_3$ ranges from $-l$ to $l$, in integer steps. 
It is necessary to introduce an arbitrary angular momentum cutoff at $l=L$.
The low partial-wave contribution for our system is then given by 

\begin{eqnarray} \label{lowsum}
2 \Gamma^{(-)}_{{\rm{L}}} &=& -\sum_{s = \pm} \; \sum_{l =
0, \frac{1}{2} , 1, \ldots}^L \; \Omega(l) \sum_{l_3 = -l}^{l}
\ln \left( \frac{ {\rm{det}}(m^2+\mathcal{H}_{(l, l_3,
s)})}{{\rm{det}}(m^2+\mathcal{H}_{(l, l_3, s)}^{{\rm{free}}})} \right) \, .
\end{eqnarray}

 The determinants can be  evaluated by using
 the Gel'fand-Yaglom method \cite{Dunne:2004sx,Dunne:2005te,Dunne:2006ac} 
 which we summarize as follows.
 Let $\mathcal{M}_1$ and $\mathcal{M}_2$ denote two second-order radial
differential operators on the interval $r \: \in \, [\, 0,\infty)$
and let $\Phi_1(r)$ and $\Phi_2(r)$ be solutions to the following
initial value problem:
\begin{equation}
\mathcal{M}_i \Phi_i(r) = 0; \quad \Phi_i(r) \sim r^{2 l} \quad {\rm{as}} \quad r \to 0 \, .
\label{initial}
\end{equation}
Then, the ratio of the determinants is given by
\begin{eqnarray}
\frac{{\rm{det}} \mathcal{M}_1}{{\rm{det}} \mathcal{M}_2} &=& \lim_{ R \to \infty} \left(  \frac{\Phi_1(R)}{\Phi_2(R)}      \right) \, .
\label{detrat}
\end{eqnarray}
Thus, taking ${\rm{det}}  \mathcal{M}_1/ {\rm{det}}  \mathcal{M}_2$ to be the determinants in (\ref{lowsum}), the calculation of the corresponding ratios
reduces to the following initial value problem:

\begin{eqnarray}
\Phi_{\pm}''(r) + \frac{4 l + 3 }{r} \Phi_{\pm}'(r) - \left( m^2 +
4 l_3 g(r) + r^2 g(r)^2 \mp [4 g(r) + r g'(r)]   \right)
\Phi_{\pm}(r) &=& 0 \, , \nonumber\\
\label{ivp}
\end{eqnarray}
with the initial value boundary condition in (\ref{initial}). The value of $\Phi$ at $r=\infty$ gives us the value of the determinant for that partial wave.
The corresponding free equation $(g(r)=0)$ is analytically soluble.

\no
It is numerically more convenient to define
\begin{eqnarray}
S_\pm^{(l, l_3 )}(r) &\equiv& \ln \left( \frac{\Phi_{l, l_3,\pm} (r)}{\Phi_{l, l_3,\pm}^{\rm free} (r)} \right) \, ,
\end{eqnarray}
and solve the corresponding initial value problem for $S_\pm(r)$, as explained in \cite{Dunne:2005te,Dunne:2007mt}.
%\begin{eqnarray} \label{Seq}
%\frac{d^2 S_{\pm}^{(l, l_3)}}{dr^2 } + \left(\frac{dS_{\pm}^{(l,
%l_3)}}{dr } \right)^2 + W_{l}(m , r) \frac{dS_{\pm}^{(l,
%l_3)}}{dr } &=& V_{l_3}(r)  \, , \nonumber \\
%S_\pm^{(l, l_3 )}(0) = 0   \quad ;   \quad   \frac{d S_\pm^{(l, l_3 )}(0) }{dr } &=& 0 \, .
%\end{eqnarray}
%We found this equation to be the most numerically stable option.
%In summary, solving this second-order non-linear equation, with the given initial value conditions and applying the Gel'fand-Yaglom theorem,  allows us to calculate 
Then, the contribution of the low-angular-momentum partial-waves to the effective action is
\begin{eqnarray}
2 \Gamma^{(-)}_{{\rm{L}}} &=& - \sum_{l = 0, \frac{1}{2}, 1 ,
\ldots}^L \; \Omega(l) \sum_{l_3 = -l}^{l} [ S_{+}^{(l,
l_3)}(\infty) + S_{-}^{(l, l_3)}(\infty) ] \, .
\label{lecont}
\end{eqnarray}
While each term in the sum over $l$ is finite and simple to compute, the sum over $l$  is divergent as $L \to \infty$.
In fact,  only after adding the high partial-wave modes and an appropriate counterterm, 
a finite and renormalized result will be obtained for the effective action.

\noindent
It remains to consider the high-mode contribution,

\begin{eqnarray} \label{gammalow}
2 \Gamma^{(-)}_{{\rm{H}}} &=& -\sum_{s = \pm \frac{1}{2}} \; \sum_{l = L
+ \frac{1}{2}}^\infty \; \Omega(l) \sum_{l_3 = -l}^{l} \ln
\left( \frac{ {\rm{det}}(m^2+\mathcal{H}_{(l, l_3, s)})}{{\rm{det}}(m^2+\mathcal{H}_{(l, l_3, s)}^{{\rm{free}}} )}
\right) \, .
\end{eqnarray}
The application of the WKB approximation to this sum has been presented in \cite{Dunne:2011fp}. Here we
quote only the final result: Using dimensional regularization ($\overline {\rm MS}$), and taking the large $L$ limit, we have 

\begin{align}
2 \Gamma^{(-)}_{{\rm{H,reg}}} &= \int_0^\infty dr \left( \frac{8 \, g(r) r^3}{3 \sqrt{\tilde{r}^2 + 4}} \right) L^2
+ \int_0^\infty dr \left( \frac{2 r^3 ( 3 \tilde{r}^3 + 8) g(r)^2}{( \tilde{r}^2 + 4)^{3/2}} \right) L \nonumber\\
&+ \int_0^\infty dr \Bigg\{ \frac{r^3}{45 ( \tilde{r}^2 +
4)^{7/2}} \Bigg[ -6 r^4 (5 \tilde{r}^4 + 28 \tilde{r}^2 + 32 )g(r)^4
\nonumber\\ &+ 15 (33 \tilde{r}^6 +  335 \tilde{r}^4  + 1192 \tilde{r}^2 + 1600) g(r)^2 \nonumber\\
&+ 10 r (15 \tilde{r}^6 + 184 \tilde{r}^4 + 776 \tilde{r}^2 + 1120) g(r)g'(r) \nonumber\\
&+ 5 r^2 (3 \tilde{r}^6 + 38 \tilde{r}^4+ 160 \tilde{r}^2 + 224) g'(r)^2 + 20 r^2 (4 + \tilde{r}^2)^2
g(r)g''(r)]
 \Bigg] \nonumber\\
&+ \frac{r^3 (20 g(r)^2 + 10 g(r)g'(r)r + g'(r)^2 r^2 ) }{12} \Bigg[
\gamma + 2  \ln L  - 2 \ln \Bigg( \frac{r}{2 + \sqrt{\tilde{r}^2 +
4}} \Bigg)
 \Bigg]
  \nonumber\\
&-  \frac{r^3 (20 g(r)^2 + 10 g(r)g'(r) r + g'(r)^2 r^2 )}{12 \,
\epsilon}\Bigg\}  + O\Bigg( \frac{1}{L} \Bigg) \, ,\nonumber\\
\label{GammaHreg}
\end{align}
where $\epsilon = - (D-4)/2$ and $\tilde{r} \equiv  r m/L$. 
Adding the appropriate UV counterterm $\delta\Gamma^{(-)}_{\rm H,reg}$, the contribution of the negative chirality sector to the effective
action becomes, still at the regularized level,

\bear
\Gamma_{\rm reg}^{(-)} = \Gamma_L^{(-)} + \Gamma^{(-)}_{{\rm{H,reg}}} + \delta \Gamma^{(-)}_{\rm H,reg} \, ,
\label{totalreg}
\ear 
where

\begin{eqnarray}
2 \delta \Gamma^{(-)}_{\rm H,reg} &=&  \Bigg( \frac{1}{\epsilon} - \gamma_E - 2
 \ln \mu  \Bigg) \Bigl(\frac{1}{24} \int_0^\infty dr \; r^3  F_{\mu \nu} F_{\mu \nu} 
+\frac{1}{16}  \int_0^\infty dr \; r^3  F_{\mu \nu} \tilde F_{\mu \nu} \Bigr)
\end{eqnarray}
and

\begin{eqnarray}\label{FF}
\frac{1}{2} \int_0^\infty dr \; r^3 (F_{\mu \nu} F_{\mu \nu})  &=&
\int_0^\infty dr \; r^3 (8 g(r)^2 + 4 g(r)g'(r) r +
g'(r)^2 r^2 ) \, ,  \\
\frac{1}{2} \int_0^\infty dr \; r^3 (F_{\mu \nu} \tilde F_{\mu \nu})  &=&
\int_0^\infty dr \; r^3 (8 g(r)^2 + 4 g(r)g'(r) r  ) \, .
\label{FFdual}
\end{eqnarray}
At this stage, the $\epsilon\to 0$ limit can be taken, and the UV divergences cancel in the sum of the last two terms on the
rhs of (\ref{totalreg}). Thus, we define

\bear
\Gamma^{(-)}_{\rm{H,ren}}(m,\mu):= \lim_{\epsilon\to 0} 
\Bigl( \Gamma^{(-)}_{{\rm{H,reg}}} + \delta \Gamma^{(-)}_{\rm H,reg}\Bigr)\;.
\label{lim}
\ear
The final, dimensionally renormalized result for the effective action becomes

\bear
\Gamma_{\rm ren} (m,\mu) = 2\Bigl(\Gamma_L^{(-)}(m) + \Gamma^{(-)}_{\rm{H,ren}}(m,\mu)\Bigr) + \Delta\Gamma_{\rm ren}(m,\mu)
\label{gammarenfin}
\ear
where $ \Delta\Gamma_{\rm ren}(m,\mu)$ is the chiral anomaly term as given in (\ref{difference}).
This expression is finite, but still contains spurious divergences in $L$ for $L\to\infty$. 
For the high-partial wave contribution, the following explicit expression was obtained in \cite{Dunne:2011fp} 
for the large $L$ behavior:

\begin{eqnarray}
 \Gamma_{\rm{H,ren}}(m,\mu)
= \int_0^\infty dr  \left( Q_{{\rm {log}}}(r)\ln L + \sum_{n = 0}^2 Q_n ( r ) L^n  +  \sum_{n = 1}^N Q_{-n}( r ) \frac{1 }{ L^n }  \right) + O\left(\frac{1}{L^{n+1}}\right) \, ,
\non
\label{gammaren}
\end{eqnarray}
%After the counterterm is added we are left 
with the following
expansion coefficients:
\begin{align}
Q_2(r) &=  \frac{8 \, g(r) r^3}{3 \sqrt{\tilde{r}^2 + 4}}\;, \nonumber \\
Q_1(r) &= \frac{2 r^3 (3 \tilde{r}^3 + 8) g(r)^2}{(
\tilde{r}^2 + 4)^{3/2}} \;,\nonumber \\
Q_{{\rm {log}}}(r) &= -\frac{1}{6}\, r^3 (20 g(r)^2 + 10 g(r)g'(r) r + g'(r)^2
r^2 ) \;,\nonumber \\
Q_0(r) &= \frac{r^3}{45 ( \tilde{r}^2 +
4)^{7/2}} \Bigg[ -6 r^4 (5 \tilde{r}^4 + 28 \tilde{r}^2 + 32 )g(r)^4
\nonumber \\ &+ 15 (33 \tilde{r}^6 +  335 \tilde{r}^4  + 1192 \tilde{r}^2 + 1600) g(r)^2  \nonumber \\
&+ 10 r (15 \tilde{r}^6 + 184 \tilde{r}^4 + 776 \tilde{r}^2 + 1120) g(r)g'(r) \nonumber \\
&+  5 r^2 (3 \tilde{r}^6 + 38 \tilde{r}^4+ 160 \tilde{r}^2 + 224) g'(r)^2 + 20 r^2 (4 + \tilde{r}^2)^2
g(r)g''(r)  \Bigg]
 \nonumber \\
&- Q_{{\rm {log}}}(r)   \ln \Bigg( \frac{ \mu r}{2 + \sqrt{\tilde{r}^2 + 4}}
\Bigg)\;,
 \nonumber \\
Q_{(-1)}(r) &= -\frac{r^3}{4 ( \tilde{r}^2 + 4)^{9/2}} \Bigg[ 6
r^4( \tilde{r}^6 + 4 \tilde{r}^4 )g(r)^4 \nonumber\\
&+2 r^2(  \tilde{r}^6 + 16 \tilde{r}^4 + 80\tilde{r}^2 +
 128)g'(r)^2 \nonumber  \\
&+ (-4 \tilde{r}^8 + 89 \tilde{r}^6 + 1104 \tilde{r}^4 + + 3456 \tilde{r}^2 + 5120 ) g(r)^2 \nonumber \\
&+16 r( 2 \tilde{r}^6 + 21 \tilde{r}^4 + 92
\tilde{r}^2 + 160)g(r) g'(r) \nonumber \\
&-4 r^2(\tilde{r}^6 + 8 \tilde{r}^4 + 16
\tilde{r}^2 )g(r) g''(r)
 \Bigg] \, .     \label{spinorQs}
\end{align}
Note that  $\Gamma_{\rm{H,ren}}(m,\mu)$ involves $L^2$, $L$ and
$\ln L$ terms  that diverge as $L \to \infty$, but these divergences must be 
exactly canceled by similar terms in the $\Gamma_{{\rm{L}}}^{(-)}(m)$ contribution.

For the class of backgrounds considered in this paper, the partial-wave-cutoff  method works well for any value of the mass up to numerical accuracy.
The effective action calculated as above is finite for any non-zero value of the mass. However, from Eq.~(\ref{spinorQs}), we note that $Q_0 (r)$ contains a term proportional 
to $\ln \mu$. Therefore, when we use on-shell (`OS') renormalization, defined by

\begin{equation}
\Gamma_{\rm ren}^{\rm OS} (m)\equiv \Gamma_{\rm ren}(m,\mu = m)
\label{defGammaOS}
\end{equation}
the leading behavior of the effective action is given by 

\begin{equation}
\Gamma_{\rm{ren}}^{\rm OS} (m) \sim \Bigg( - \int_0^\infty dr \, Q_{{\rm {log}}}(r) \Bigg) \ln m \, , \quad \quad   m \to 0 \, .
\end{equation}
Therefore, in the small $m$ regime, it is convenient to  introduce a modified effective action defined as \cite{Dunne:2011fp}
\begin{equation}
\tilde{\Gamma}_{\rm{ren}} (m) \equiv \Gamma_{\rm{ren}} (m,\mu) + \Bigg(  \int_0^\infty dr \, Q_{{\rm {log}}}(r) \Bigg) \ln \mu \quad
\Bigl( =  \Gamma_{\rm ren}(m,\mu =1) \Bigr)
 \,,
\label{defGammatilde}
\end{equation}
which is independent of the renormalization scale $\mu$, and 
finite at $m=0$. Here, we go beyond the findings of Ref.~\cite{Dunne:2011fp} by including results  for $\tilde{\Gamma}_{\rm ren} (m)$
for the strictly massless case. 

For the large-mass regime, we can compare our results against the large-mass expansion. This is better achieved if we work with the 
physically renormalized effective action $\Gamma_{\rm ren}^{\rm OS}(m)$, which remains finite as $m \to \infty$.
We improve on the large-mass results shown in \cite{Dunne:2011fp} 
(that correspond to $\tilde{\Gamma}_{\rm ren}(m)$) by analyzing $\Gamma_{\rm ren}^{\rm OS}(m)$ and explicitly comparing leading and subleading large-mass expansion coefficients obtained  numerically with the partial-wave-cutoff method against their exact analytical values.

\section{Perturbative results}
\label{perturbative}
\renewcommand{\theequation}{4.\arabic{equation}}
\setcounter{equation}{0}

Before starting on our numerical analysis of the effective action, which will be intrinsically nonperturbative,
in this section we perform a number of perturbative computations that will help us to interpret those
nonperturbative results, as well as to verify their numerical accuracy. In these computations we use
the worldline formalism along the lines of \cite{strassler,5,41}, and as it is usual in that formalism, as a byproduct of our spinor QED calculations we will obtain also the corresponding quantities for Scalar QED. The latter will be included
here, for their own interest as well as with a view on future extensions of this work to the Scalar QED case
(Scalar QED quantities will be given a subscript `$\rm scal$').

\subsection{Large mass expansion of the effective action}
\label{hk}

In this subsection we calculate the leading and subleading terms in the inverse mass (= heat kernel) expansion of the
one-loop scalar and spinor QED effective actions, following the approach of \cite{5,25}. The starting point is Feynman's
worldline path integral representation of the scalar loop effective action \cite{feynman:pr80,41}
(note that in our present conventions the effective action is defined with the opposite sign relative to the conventions of \cite{41}),

\begin{equation}
\Gamma_{\rm scal}[A]=
-\int^\infty_0 \frac{dT}{T} e^{-m^2 T}\int \mathcal{D}x(\tau)e^{-\int^T_0 d\tau\dot{x}^{2}/4-ie\int^T_0 d\tau \dot{x}^{\mu}A_{\mu}(x)} \;.
\label{wlint}
\end{equation}
Here, at fixed proper-time $T$, the path integral runs over all closed loops in spacetime with periodicity $T$. 
We will generally Taylor expand the Maxwell field $A_\mu(x)$ at the loop center-of-mass $x_0$, defined by

\bear
x_0^{\mu} &\equiv& \frac{1}{T}\int_0^Td\tau x^{\mu}(\tau)\;,
\label{defxo}
\ear
and then use Fock-Schwinger gauge to write the coefficients of this expansion in terms of the
field strength tensor $F_{\mu\nu}(x_0)$ and its derivatives \cite{5}. The first few terms in this expansion are

\begin{equation}
 A_{\mu}(x=x_{0}+y)= -\frac{1}{2} F_{\mu\nu}(x_{0})y^{\nu}-\frac{1}{3} F_{\mu\nu,\alpha}(x_{0})y^{\nu}y^{\alpha}
 -\frac{1}{8}F_{\mu\nu,\alpha\beta}(x_{0})y^{\nu}y^{\alpha}y^{\beta} + \cdots
 \label{fs}
\end{equation}
Combining the expansion of the interaction exponential in the path integral (\ref{wlint}) with the Fock-Schwinger expansion (\ref{fs}),
the path integration can be reduced to gaussian form. So, its performance requires only the knowledge of the free path integral
normalization factor, and the two-point correlator. Those are, respectively,  

\begin{eqnarray}
\int\mathcal{D}y~{\rm exp}\left[-\int ^{T}_{0}d\tau\frac{1}{4}\dot{y}^{2}\right]=(4\pi T)^{-D/2}\;,
\label{freepi}
\end{eqnarray}
and

\begin{eqnarray}
\langle y^{\mu}(\tau) y^{\nu}(\tau') \rangle = - \delta^{\mu\nu}G_B(\tau,\tau')\;,
\label{wc}
\end{eqnarray}
with the worldline Green's function

\bear
G_B(\tau,\tau') = \mid\tau-\tau'\mid - \frac{(\tau-\tau')^2}{T}\;.
\label{DefGB}
\ear

One then collects the terms with a fixed power of $T$,  
and obtains the inverse mass expansion of the effective action in the form

\begin{equation}
\Gamma_{\rm scal}[F] = \int_0^\infty \!{dT\over T} \; 
\frac{{\rm e}^{-m^2 T}}{(4\pi T)^{D/2}} \; 
{\rm tr} \; \int \! dx_0 \; \sum_{n=1}^N \; 
\frac{(-T)^n}{n!} \; O_n[F] \,,
\end{equation}\no
where $O_n(F)$ contains the operators in the effective action of mass dimension $2n$. For the physically renormalized
effective action $\Gamma^{\rm OS}_{\rm scal,spin}$, the lowest non-vanishing mass level is $n=3$, which therefore
dominates in the large mass limit. In the following, we calculate this leading contribution and also the
subleading $n=4$ terms, for both scalar and spinor QED. Since we work on the finite part of the effective action only,
we can set $D=4$ from now onward.

Starting with the leading order, this is given by
the term in the effective action involving two copies of the second
term in the expansion (\ref{fs}).  Denoting this term by $\Gamma^{\partial F \partial F}$,
we have (in an obvious notation and omitting the argument $x_0$ of the field strength tensors)

\begin{eqnarray}
 \Gamma^{\partial F \partial F}[A]&=&-\frac{(-i)^{2}}{2!}\int^{\infty}_0\frac{dT}{T}e^{-m^{2}T}\int d^4x_{0}\int^T_0d\tau_{1}\int^T_0d\tau_{2}\nonumber \\
&&\times \int \mathcal{D}y\left[\frac{1}{9}~\dot{y}^{\mu_{1}}F_{\mu_{1}\nu_{1},\alpha_{1}}y^{\nu_{1}}y^{\alpha_{1}}
 \dot{y}^{\mu_{2}}F_{\mu_{2}\nu_{2},\alpha_{2}}y^{\nu_{2}}y^{\alpha_{2}}~\right]e^{-\int^{T}_{0}d\tau\dot{y}^{2}/4}
\nonumber\\
&=& 
\frac{1}{18}
\int^{\infty}_0\frac{dT}{T}e^{-m^{2}T}
\frac{1}{(4\pi T)^2}
\int d^4x_{0}\int^T_0d\tau_{1}\int^T_0d\tau_{2}\,
F_{\mu_{1}\nu_{1},\alpha_{1}}  F_{\mu_{2}\nu_{2},\alpha_{2}} {\cal M}\;,
\nonumber\\
\label{Gammaleading}
 \end{eqnarray}
where we have set $e=1$ as usual, and 

\begin{eqnarray}
\mathcal{M}&=&\ddot{G}_{B}(\tau_{1},\tau_{2})G^{2}_{B}(\tau_{1},\tau_{2})\Big\{\delta^{\mu_{1}\mu_{2}}\delta^{\nu_{1}\nu_{2}}\delta^{\alpha_{1}\alpha_{2}}+\delta^{\mu_{1}\mu_{2}}\delta^{\nu_{1}\alpha_{2}}\delta^{\alpha_{1}\nu_{2}}\Big\}\nonumber\\
&&+\dot{G}^{2}_{B}(\tau_{1},\tau_{2})G_{B}(\tau_{1},\tau_{2})\Big\{\delta^{\mu_{1}\nu_{2}}\delta^{\nu_{1}\mu_{2}}\delta^{\alpha_{1}\alpha_{2}}+\delta^{\mu_{1}\nu_{2}}\delta^{\nu_{1}\alpha_{2}}\delta^{\alpha_{1}\mu_{2}}\nonumber \\
&&\hspace{110pt}+\delta^{\mu_{1}\alpha_{2}}\delta^{\nu_{1}\mu_{2}}\delta^{\alpha_{1}\nu_{2}}+
\delta^{\mu_{1}\alpha_{2}}\delta^{\nu_{1}\nu_{2}}\delta^{\alpha_{1}\mu_{2}}\Big\}\nonumber\\
&\equiv&\ddot{G}_{B}(\tau_{1},\tau_{2})G^{2}_{B}(\tau_{1},\tau_{2})\delta_{1}+\dot{G}^{2}_{B}(\tau_{1},\tau_{2})G_{B}(\tau_{1},\tau_{2})\delta_{2}\;.
\end{eqnarray}
Through an integration-by-parts, we can replace $\ddot{G}_{B}(\tau_{1},\tau_{2})G^{2}_{B}(\tau_{1},\tau_{2})$ by 
$-2\dot{G}^{2}_{B}(\tau_{1},\tau_{2})G_{B}(\tau_{1},\tau_{2})$ in the first term of $\cal M$. Next, we use the
Bianchi identity to show that 

\bear
F_{\mu_{1}\nu_{1},\alpha_{1}}  F_{\mu_{2}\nu_{2},\alpha_{2}}\delta_1 = 
- F_{\mu_{1}\nu_{1},\alpha_{1}}  F_{\mu_{2}\nu_{2},\alpha_{2}}\delta_2 = \frac{3}{2} F_{\mu\nu,\alpha}^2\;.
\label{usebianchi}
\ear
Thus,

\bear
F_{\mu_{1}\nu_{1},\alpha_{1}}  F_{\mu_{2}\nu_{2},\alpha_{2}} {\cal M}
= -\frac{9}{2}\dot{G}^{2}_{B}(\tau_{1},\tau_{2})G_{B}(\tau_{1},\tau_{2}) F_{\mu\nu,\alpha}^2\;.
\label{elM}
\ear
Next, we perform the $\tau_i$ integrals. Here, as usual, one can use the unbroken reparametrization invariance to 
set $\tau_2=0$ and rescale $\tau_1= Tu$, with the result

\bear
\int_0^Td\tau_1 \int_0^Td\tau_2 \,\dot{G}^{2}_{B}(\tau_{1},\tau_{2})G_{B}(\tau_{1},\tau_{2})
= T^2 \int_0^1 du(1-2u)^2u(1-u) = \frac{T^2}{30}\;.
\label{intdu}
\ear
Performing the final $T$ - integration and putting things together we get our final result,

 \begin{equation}
   \Gamma_{\rm scal}^{\partial F\partial F}[A]=-\frac{1}{1920\pi^{2}m^{2}}\int d^{4}x_{0}~ F^{2}_{\mu_{2}\nu_{2},\alpha_{2}}(x_{0})  \;.
   \label{Gammascalleadfin}
   \end{equation}

To get the corresponding result for the spinor QED case, we make use of the ``Bern-Kosower replacement rule'' \cite{berkos},
according to which the result for the spinor loop is inferred from the scalar result --in the integrand after the integration-by-parts-- by replacing 

\bear
\dot{G}^{2}_{B}(\tau_{1},\tau_{2})\to \dot{G}^{2}_{B}(\tau_{1},\tau_{2})- G^{2}_{F}(\tau_{1},\tau_{2})\;,
\label{rrule}
\ear
with the ``fermionic'' worldline Green's function $G_F(\tau,\tau') = {\rm sign}(\tau-\tau')$. This changes the integral (\ref{intdu}) into

\bear
\int_0^Td\tau_1 \int_0^Td\tau_2 \,\bigl[\dot{G}^{2}_{B}(\tau_{1},\tau_{2})-G^{2}_{F}(\tau_{1},\tau_{2})\bigr] G_{B}(\tau_{1},\tau_{2})
%&=& T^2 \int_0^1 du\bigl[(1-2u)^2-1\bigr] u(1-u) 
%\nonumber\\
&=& -\frac{2}{15}T^2\;.\nonumber\\
\label{intduspin}
\ear
Also, the global normalization has to be changed by a factor of $-2$. Thus, our result for the leading term in the spinor QED large mass
expansion is

 \begin{equation}
\Gamma^{\partial F\partial F}[A]= 8  \Gamma^{\partial F\partial F}_{\rm scal}[A]
   = -\frac{1}{240\pi^{2}m^{2}}\int d^{4}x_{0}~ F^{2}_{\mu_{2}\nu_{2},\alpha_{2}}(x_{0})  \;.
\label{Gammaspinleadfin}
\end{equation}
For the spinor QED case,  this term was also computed in \cite{lepash}.   

It should be noted that, at the same mass level, there could have been a contribution involving the product of the first and third terms
in the Fock-Schwinger expansion (\ref{fs}). However, it drops out due to the vanishing of the coincidence
limits $G_B(\tau,\tau) = \dot G_B(\tau,\tau) = 0$.

 At the next mass level, which is of mass dimension eight, one again finds that the non-vanishing contributions
 with only two fields involve two copies of the third term in the Fock-Schwinger expansion (\ref{fs}). The computation
 is analogous to the previous one, including the use of the ``replacement rule''. Here we only quote the result:
 
  \begin{eqnarray}
  \Gamma_{\rm scal}^{\partial\partial F\partial \partial F}[A]&=&\frac{1}{26880\pi^2m^4}\int d^{4}x_{0}F_{\mu_{2}\nu_{2},\alpha_{2}\beta_{2}}^{2}\;,
  \label{ddFscal}\\
   \Gamma^{\partial\partial F\partial \partial F}[A]  &=&\frac{1}{2240\pi^2m^4}\int d^{4}x_{0}F_{\mu_{2}\nu_{2},\alpha_{2}\beta_{2}}^{2}\;.
   \label{ddFspin}
      \end{eqnarray}
  
At this same subleading mass dimension level, we also have the terms with four $F$'s. Their contributions are contained in the 
effective Lagrangians for a constant field, due to Heisenberg and Euler \cite{Heisenberg:1935qt}  in the
spinor QED case, and Weisskopf \cite{weisskopf} for the scalar QED case.
From those Lagrangians, one easily finds

\begin{eqnarray}
\Gamma_{\rm scal}^{FFFF}[A] &=&- \frac{1}{16\pi^2 m^4}\int d^4x_0 
\Bigl[\frac{1}{360}{\rm tr} (F^4) + \frac{1}{288} ({\rm tr} F^2)^2\Bigr]\;,
\label{quartscal}\\
\Gamma^{FFFF}[A]&=&- \frac{1}{16\pi^2 m^4}\int d^4x_0 
\Bigl[\frac{7}{90}{\rm tr} (F^4) - \frac{1}{36} ({\rm tr} F^2)^2\Bigr]\;.
\end{eqnarray}
Finally, we insert our background field, defined by  (\ref{deffield}) and (\ref{g2}) (with $\rho=\nu =1$) and expand in inverse powers of mass. 
In the limit when $\alpha\to 0$, the coefficients of the inverse squared and inverse quartic terms, up to cubic order in $\alpha$, are
respectively given by

\begin{eqnarray}
c_{{\rm scal},2}&=& - \frac{2}{15} \alpha ^3 \left( \ln (2 \alpha )+ \gamma_E +\frac{19}{80}\right)
-\frac{31 \alpha ^2}{600}+\frac{23 \alpha }{1200}-\frac{1}{75}\;, \nonumber\\
c_{{\rm scal},4}&=&
\frac{203}{270} \alpha ^3 \left( \ln (4 \alpha )+ \gamma_E -\frac{60601}{59682}\right)+\frac{11}{60} \alpha ^2 \left( \ln (4 \alpha )+ \gamma_E +\frac{32663}{776160}\right)+\frac{2941 \alpha }{66150}-\frac{107 }{105840}\;,\nonumber\\
\label{cscal24}
\end{eqnarray}
in the scalar case, whereas for the spinor case,

\begin{eqnarray}
c_{{\rm spin},2}&=& 8\, c_{{\rm scal},2}, \nonumber\\
c_{{\rm spin},4}&=& \frac{232}{135} \alpha ^3 \left( \ln (4 \alpha )+ \gamma_E +\frac{137}{588}\right)+\frac{2}{15} \alpha ^2 \left( \ln (4 \alpha )+ \gamma_E +\frac{8819}{35280}\right)+\frac{3149 \alpha }{33075}+\frac{683 }{13230} \;,\nonumber\\
\label{cspin24}
\end{eqnarray}
where $\gamma_E\simeq 0.57721$ is the Euler-Mascheroni constant. These expressions were obtained with MATHEMATICA~\cite{math}.

\subsection{Two-point functions}
\label{2point}

We will now compute the two-point contributions to the scalar and spinor effective actions in the background 
defined by (\ref{deffield}) and (\ref{g2}). We will
set $\nu = \rho =1$, but keep $\alpha$, $m$ and $\mu$ general, at least initially.

For the scalar case, to get the two-point contribution we start again from the worldline representation of the effective action (\ref{wlint}),
and expand out the interaction exponential to second order. This yields 

\bear
\Gamma^{(2)}_{\rm scal}= \frac{1}{2} \int_0^{\infty}\frac{dT}{T}e^{-m^2T}\int_0^Td\tau_1d\tau_2\int
\mathcal{D}x\prod_{i=1}^2\dot{x}_i\cdot A(x_i)\,{\rm exp}\Big\lbrack-\int_0^Td\tau\frac{\dot{x}^2}{4}\Big\rbrack
\, .
\label{Gam}
\ear
where we have put $x_i \equiv x(\tau_i)$. Fourier transforming $A_\mu(x)$,
 
\bear
A_\mu(x)=\int \frac{d^Dk}{(2\pi)^D}e^{ik\cdot x}\bar{A}_\mu(k)
\label{four}
\ear
we find that

\bear
\bar{A}_\mu(k)=-i\eta^3_{\mu\nu}k^{\nu}\bar{b}(k^2,\alpha)
\ear
where $\bar b(k^2,\alpha)$ can be written as

\bear
\bar b(k^2,\alpha) = \frac{\pi^2}{2} \,\e^{\alpha}\, \Gamma(-2,\alpha;k^2/4)
\label{b}
\ear
with $\Gamma$ the  \emph{generalized incomplete gamma function}

\bear
\Gamma(a,x;b)=\int_x^\infty dz\, z^{a-1}e^{-z-bz^{-1}} \, .
\ear
Introducing (fictitious) polarization vectors by

\bear
\varepsilon_{i\mu}&:=&\eta^3_{\mu\nu}k_i^\nu~~~,~~~i=1, 2, \label{defeps}
\ear
and the two-point function in momentum space,

\bear
\Gamma^{(2)}_{\rm scal}[k_1,\varepsilon_1;k_2,\varepsilon_2]&=&- \int_0^\infty\frac{dT}{T}e^{-m^2T}\int_0^T d\tau_1d\tau_2\int Dx
\prod_{i=1}^2\varepsilon_i\cdot\dot{x}_i\,e^{ik\cdot x}e^{-\int_0^Td\tau\frac{\dot{x}^2}{4}}\non
\ear
we can then rewrite $\Gamma^{(2)}_{\rm scal}$ as

\bear
\Gamma^{(2)}_{\rm scal}= \frac{1}{2} \prod_{i=1}^2\int \frac{d^Dk_i}{(2\pi)^D}\bar{b}(k_i^2,\alpha)\Gamma^{(2)}_{\rm scal}[k_1,\varepsilon_1;k_2,\varepsilon_2]
\, .
\label{gamma}
\ear
The calculation of $\Gamma^{(2)}_{\rm scal}[k_1,\varepsilon_1;k_2,\varepsilon_2]$ is a standard textbook calculation, and we give the
result here only. 
In the $\overline{\rm MS}$ scheme with mass scale $\mu$, one finds

\bear
\Gamma^{(2)}_{\rm scal,ren}[k_1,\varepsilon_1;k_2,\varepsilon_2] &=&    (2\pi)^D\delta(k_1+k_2) 
( \varepsilon_1\cdot \varepsilon_2 k_1\cdot k_2-\varepsilon_1\cdot k_2 \varepsilon_2\cdot k_1)
\Pi_{\rm scal,ren}(k_1^2,m,\mu), \nonumber\\
\Pi_{\rm scal,ren}(k^2,m,\mu) &=&- \frac{1}{(4\pi)^2}
\Bigg\{\frac{1}{3}\ln \frac{m^2}{\mu^2}
-\frac{8}{9}\Bigl(1+3\frac{m^2}{k^2}\Bigr)
+\frac{1}{3}\Bigl(1+4\frac{m^2}{k^2}\Bigr)^{\frac{3}{2}}\,{\rm ArcTanh}\bigg\lbrack\frac{\sqrt{1+4\frac{m^2}{k^2}}}{1+2\frac{m^2}{k^2}}\bigg\rbrack\Bigg\}
\, .
\non
\label{Gamma2scalk}
\ear
Setting now $k=k_1= - k_2$ and using (\ref{defeps}) together with $\eta^{3T} = - \eta^3$ and $(\eta^3)^2= - \Eins$, we compute

\bear
( \varepsilon_1\cdot \varepsilon_2 k_1\cdot k_2-\varepsilon_1\cdot k_2 \varepsilon_2\cdot k_1)
=  k^4 \, .
\ear
After using the $\delta$ - function in (\ref{Gamma2scalk}) to remove the $k_2$ - integral, 
our final renormalized result for the two-point function becomes

\bear
\Gamma^{(2)}_{\rm scal,ren}[m,\mu]= \frac{1}{2}\int \frac{d^4k}{(2\pi)^4}k^4 \bar{b}^2(k^2,\alpha) \Pi_{\rm scal}(k^2,m,\mu)
\, .
\label{gammafinal}
\ear
We will also need its massless limit. Setting now $\mu=1$ and taking the limit $m\to 0$, we find 

\bear
\Pi_{\rm scal,ren}(k^2,0,1) = \frac{1}{(4\pi)^2}
\Bigl( \frac{8}{9} - \frac{1}{3} \ln k^2\Bigr)
\, .
\label{Piscalmzero}
\ear
Even in the massless case, it seems not to be possible to evaluate the triple integral in (\ref{gammafinal}) in closed form.
However, it is not difficult to determine its asymptotic behavior for $\alpha \to 0$. We find

\bear
\Gamma^{(2)}_{\rm scal, ren}[0,1] = \frac{1}{48}(\ln \alpha)^2 + \Bigl(-\frac{23}{288} + \frac{1}{8} \ln 2 - \frac{1}{24}\gamma_E \Bigr) \ln \alpha + \, {\rm finite}
\, .
\non
\label{asympscal}
\ear

Moving on to the spinor case, for $\Gamma^{(2)}_{\rm spin,ren}[m,\mu]$ we get the same formula (\ref{gammafinal}) with
$\Pi_{\rm scal,ren}$ replaced by the spinor QED vacuum polarization $\Pi_{\rm ren}$,

\bear
\Pi_{\rm ren}(k^2,m,\mu) &=&- \frac{1}{(4\pi)^2}
\Bigg\{\frac{4}{3}\ln \frac{m^2}{\mu^2}
- \frac{20}{9}\Bigl(1-\frac{12}{5}\frac{m^2}{k^2}\Bigr)
+\frac{8}{3}\Bigl(1-2\frac{m^2}{k^2}\Bigr)\sqrt{1+4\frac{m^2}{k^2}}
{\rm ArcCoth} \sqrt{1+4\frac{m^2}{k^2}}\Bigg\}
\, .
\non
\label{Pispin}
\ear
In the massless limit, this yields

\bear
\Pi_{\rm ren}(k^2,0,1) = \frac{1}{(4\pi)^2}
\Bigl(\frac{20}{9} - \frac{4}{3} \ln k^2\Bigr)
\, ,
\label{Pispinmzero}
\ear
and for the small $\alpha$ limit, we find

\bear
\Gamma^{(2)}_{\rm ren}[0,1] =  \frac{1}{12}(\ln \alpha)^2 + \Bigl(-\frac{11}{72} + \frac{1}{2}\ln 2 - \frac{1}{6}\gamma_E \Bigr) \ln \alpha + \, {\rm finite}
\, .
\non
\label{asympspin}
\ear
We remark that the leading $(\ln \alpha)^2$  terms in (\ref{asympscal}), (\ref{asympspin}) come from the $\ln k^2$ terms in (\ref{Piscalmzero}), (\ref{Pispinmzero}), so that
their coefficients are related to the $\frac{1}{\epsilon}$ poles of the two-point functions, and thus ultimately to the QED $\beta$ - functions.

\subsection{Finiteness of the quartic and higher contributions for $m=\alpha=0$}
\label{fryconst}

Next, we consider the quartic contribution of
 the one-loop effective action. Contrary to the case of the
 two-point function treated in the previous subsection, at the four-point level a detailed calculation is
 out of the question, and our only goal is to demonstrate that the four-point function is
 finite in the double limit $m, \alpha \to 0$. Thus here we consider only the $\alpha = 0$ case, 
 and wish to show that there are no divergences in the zero mass limit.
Since the four-point contribution is already UV finite, contrary to the case of the
two-point function here we can also set $D=4$ from the beginning. 

We start with the scalar QED case. Expanding the worldline path integral (\ref{wlint}) to quartic order,
we can write this quartic contribution to the effective
action as

\begin{eqnarray}
\Gamma_{\rm scal}^{(4)}[A] &=& 
- \frac{1}{4!}
\int_0^{\infty}
{dT\over T} \, {\rm e}^{-m^2T}
\int_0^Td\tau_1d\tau_2 d\tau_3 d\tau_4
\int {\cal D}x\, 
\prod_{i=1}^4
\dot x_i\cdot A(x_i)
\,
{\rm exp} \left[ 
- \int_0^T \!\!\! d\tau  \frac{\dot x^2}{4} 
\right]\;.\non
\label{Gamma4}
\end{eqnarray}
\no
As in the two-point case above, we next
Fourier transform $A_\mu(x)$, where due to our setting $\alpha =0$  the Fourier
transform can now be given more explicitly in terms of the modified Bessel function of the second kind
$K_2(x)$:

\begin{eqnarray}
\bar A_{\mu}(k) &=& -i \eta^3_{\mu\nu}k^{\nu}\bar a(k^2),
\end{eqnarray}

\begin{eqnarray}
\bar a(k^2) &=& \bar b(k^2,0) = 4 \pi^2 \frac{K_2(\sqrt{k^2})}{k^2}\, .
\label{a}
\end{eqnarray}
Later on, we will need the small and large $k$ behavior of $\bar a(k^2)$,

\begin{eqnarray}
\bar a(k^2) &=& \frac{8\pi^2}{k^4} - \frac{2 \pi^2}{k^2} + \cdots \label{asmall}\\
\bar a(k^2) &=& 2\sqrt{2}\pi^{5/2}\frac{1}{k^{5/2}}\e^{-\sqrt{k^2}} + \cdots \label{slarge}
\end{eqnarray}
Introducing polarization vectors as in (\ref{defeps}), we can rewrite (\ref{Gamma4}) as 

\begin{eqnarray}
\Gamma^{(4)}[A] &=& 
-\frac{1}{4!}
\prod_{i=1}^4
\int \frac{d^4k_i}{(2\pi)^4}\bar a(k_i^2)
\int_0^{\infty}
{dT\over T} \, {\rm e}^{-m^2T}
\int_0^Td\tau_1d\tau_2 d\tau_3 d\tau_4
\nonumber\\
&&\times
\int {\cal D}x\, 
\prod_{i=1}^4
\varepsilon_i\cdot \dot x_i \,\e^{ik_i\cdot x_i}
{\rm exp} \left[ 
- \int_0^T \!\!\! d\tau  \frac{\dot x^2}{4} 
\right]\;.\non
\label{Gamma4k}
\end{eqnarray}
\no
We separate off the zero mode $x_0$ contained in the path integral,

\begin{eqnarray}
{\dps\int}{\cal D}x &=&
{\dps\int}d x_0{\dps\int}{\cal D} y\;,\nonumber\\
x^{\mu}(\tau) &=& x^{\mu}_0  +  
y^{\mu} (\tau )\;,\nonumber\\
\int_0^T d\tau\,   y^{\mu} (\tau ) &=& 0\;.\nonumber\\
\label{split}
\end{eqnarray}
Its integral gives the usual $\delta$ - function for energy-momentum conservation. 
Thus we have 

\begin{eqnarray}
\Gamma^{(4)}[A] = -\frac{1}{4!}\prod_{i=1}^4
\int \frac{d^4k_i}{(2\pi)^4}\bar a(k_i^2)
(2\pi)^4\delta^4(\sum k_i)
\Gamma[k_1,\varepsilon_1;\cdots;k_4,\varepsilon_4]\;,
\label{Gamma4fin}
\end{eqnarray}
where $\Gamma$ is the worldline representation of the 
off-shell Euclidean four-photon amplitude in momentum space:

\begin{eqnarray}
\Gamma[k_1,\varepsilon_1;\cdots;k_4,\varepsilon_4]
&=&
-
\int_0^{\infty}
{dT\over T} \, {\rm e}^{-m^2T}
\int_0^Td\tau_1d\tau_2 d\tau_3 d\tau_4
\nonumber\\
&&\times
\int {\cal D}y\, 
\prod_{i=1}^4
\varepsilon_i\cdot \dot y_i \,\e^{ik_i\cdot y_i}
{\rm exp} \left[ 
- \int_0^T \!\!\! d\tau  \frac{\dot y^2}{4} 
\right]\;.\non
\label{4phot}
\end{eqnarray}
\no
After performing the path integral, suitable integrations by parts, a rescaling $\tau_i = Tu_i,i=1,\ldots,4$
and performance of the global $T$ integral, one obtains (see \cite{41} for the details) 

\begin{eqnarray}
\Gamma[k_1,\varepsilon_1;\cdots;k_4,\varepsilon_4]
&=&
-\frac{1}{(4\pi)^2} 
\int_0^1 du_1du_2 du_3 du_4
\frac{Q_4(\dot G_{B12},\ldots,\dot G_{B34})}
{\Bigl(m^2 -\frac{1}{2} \sum_{i,j=1}^4 G_{Bij}k_i\cdot k_j\Bigr)^2}\;.
\label{4photfin}\non
\end{eqnarray}
\no
Here, 
$G_{Bij} \equiv G_{B}(u_{i},u_{j})= |u_i-u_j|- (u_i-u_j)^{2}$
is the worldline Green's function and $\dot G_{Bij} = {\rm sign}(u_i-u_j) - 2(u_i-u_j)$
its derivative. $Q_4$ is a polynomial in the various $\dot G_{Bij}$'s, as well as in the
momenta and polarizations. 

Now, the QED Ward identity implies that (\ref{4photfin}) is $O(k_i)$ in each of the four momenta, which can
also be easily verified using properties of the numerator polynomial $Q_4$ given in \cite{41}. Using this fact
and (\ref{asmall}) in (\ref{Gamma4k}), we see that there is no singularity at  $k_i=0$, and convergence at large $k_i$
is assured by (\ref{slarge}). Further, after specializing the $u_i$ integrals to the standard ordering
$u_1\geq u_2\geq u_3\geq u_4=0$ (all ordered sectors give the same here by permutation symmetry) 
and changing from the $u_i$ variables to standard Schwinger parameters $a_i$, in the denominator,
we find the standard off-shell four point expression 

\bear
-\frac{1}{2} \sum_{i,j=1}^4 G_{Bij}k_i\cdot k_j = 
a_1a_3(k_1+k_2)^2 + a_2a_4(k_2+k_3)^2 + a_1a_2k_1^2 + a_2a_3 k_2^2 + a_3 a_4 k_3^2 + a_4a_1k_4^2\;.
\non
\label{denom}
\ear
Thus, for nonzero mass the denominator in the rhs of (\ref{4photfin})  is alway positive for our Euclidean momenta. Taking now the massless
limit, any divergence of the effective action in this limit would have to come from a non-integrable singularity
due to a zero of (\ref{denom}). However, it is easily seen that all such zeros require multiple pinches 
in the total $k_1,\ldots , a_4$ space, whose measure factors render integrable the corresponding second-order pole.

This method can be easily extended to show that also all higher $N$ - point functions are finite in the double limit $m, \alpha \to 0$.

\section{Nonperturbative results}
\label{nonperturbative}
\renewcommand{\theequation}{5.\arabic{equation}}
\setcounter{equation}{0}

We now show our numerical results for the (Spinor) QED effective action in the family of backgrounds (\ref{g2}).
In the following we use $\rho =1 $ throughout and, unless stated otherwise, also $\nu =1$.

%%%%%%%%%%%%%%%%%%%%%%%%%%%%%%%%%%%%%%%%%%%%%%%%%%%%%%%%%%%%%%%%%%%%%%%%%%%%%%%%%%%%%%%%%%%%%%%%%%%%%%%%%%%%%%%%%%%%%%%%%

\begin{figure}[t]
\begin{center}
\includegraphics[width=0.45\textwidth]{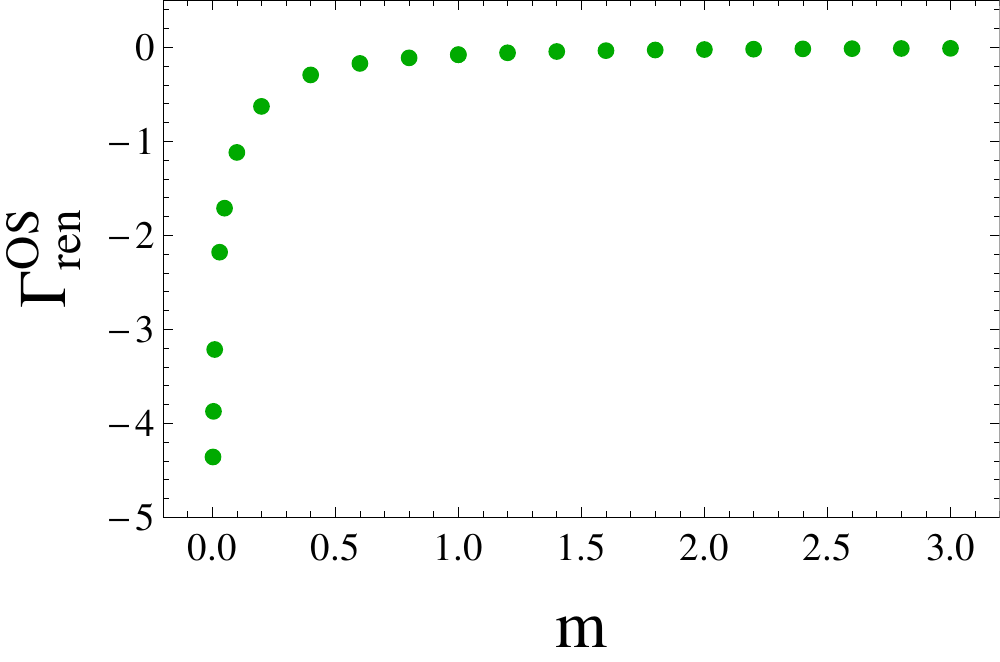}
\end{center}
\caption{The effective action  $\Gamma_{{\rm ren}}^{\rm OS}(m)$ for  $\alpha=1/100$. }
\label{fig1}

\end{figure}

\begin{figure}[t]
\begin{center}
\includegraphics[width=0.45\textwidth]{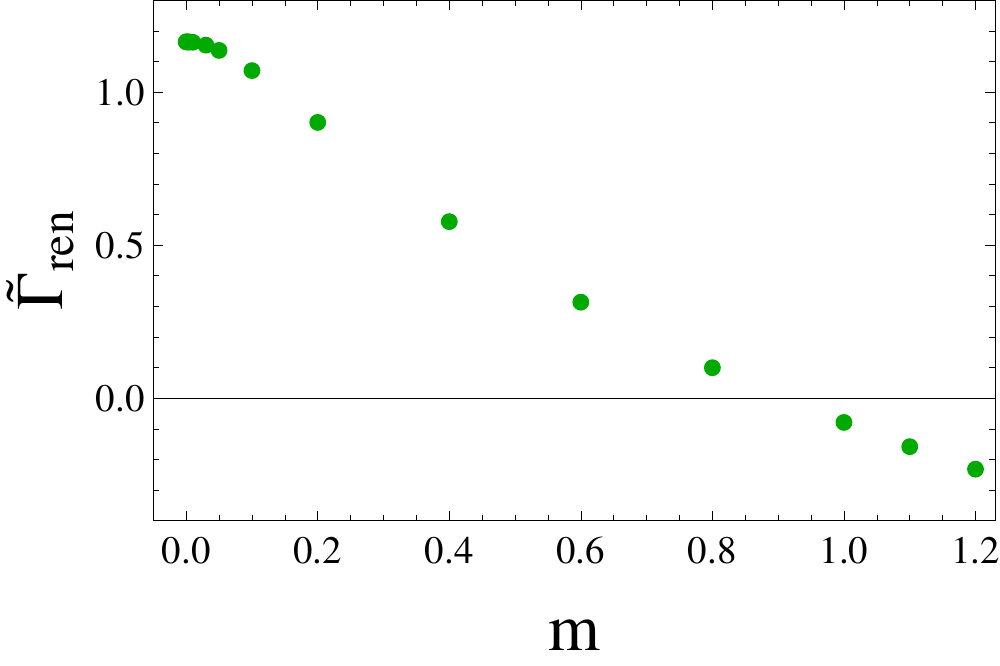}
\end{center}
\caption{The effective action  $\tilde\Gamma_{{\rm ren}}(m)$ for  $\alpha=1/100$. }
\label{fig2}

\end{figure}

We calculated the physically renormalized effective action for the full mass range. Figure~\ref{fig1} shows the behavior of  $\Gamma_{{\rm ren}}^{\rm OS}(m)$  for $\alpha=1/100$. 
%
%
%
% PLOT
%
The effective action has the expected behavior in the small-mass regime;  the leading term is proportional to
$(\int F^2 dx) {\rm ln}\, m$. After removing this term, the new effective action, $\tilde\Gamma_{\rm ren}(m)$, is 
finite as $m \to 0$, but divergent for $m\to \infty$. A plot is shown in Fig.~\ref{fig2}.

\begin{figure}[t]
\begin{center}
\includegraphics[width=0.45\textwidth]{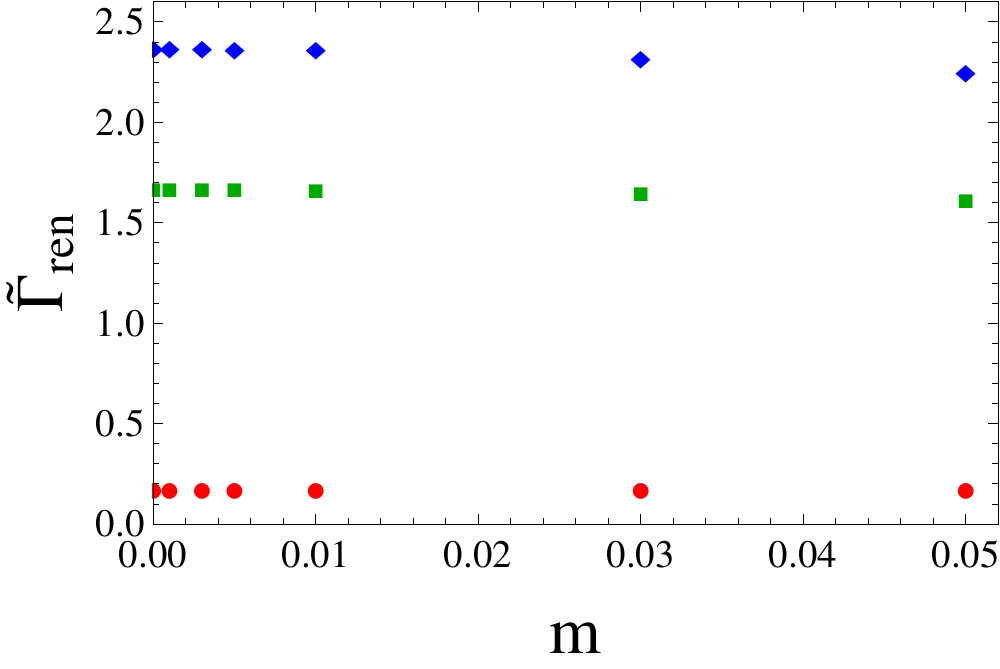}
\end{center}
\caption{The effective action $\tilde\Gamma_{\rm ren}(m)$ in the small-mass regime for different values of 
$\alpha$. Dots correspond to $\alpha=1/10$, squares to $\alpha=1/200$ and diamonds to $\alpha=1/450$.}
\label{fig3}
\end{figure}

The finiteness of  $\tilde\Gamma_{\rm ren}(m)$ for $m\to 0$ was already shown in  \cite{Dunne:2011fp}, and  for the scalar QED case, a comparison
was made with the leading term of the derivative expansion, finding good agreement.
However, going beyond the results of \cite{Dunne:2011fp} here we have also obtained extensive results for small masses and, 
in addition, we have calculated the effective action taking the 
mass to be exactly zero. Results are shown if Fig.~\ref{fig3} for different vales of $\alpha$.

\begin{figure}[t]
\begin{center}
\includegraphics[width=0.45\textwidth]{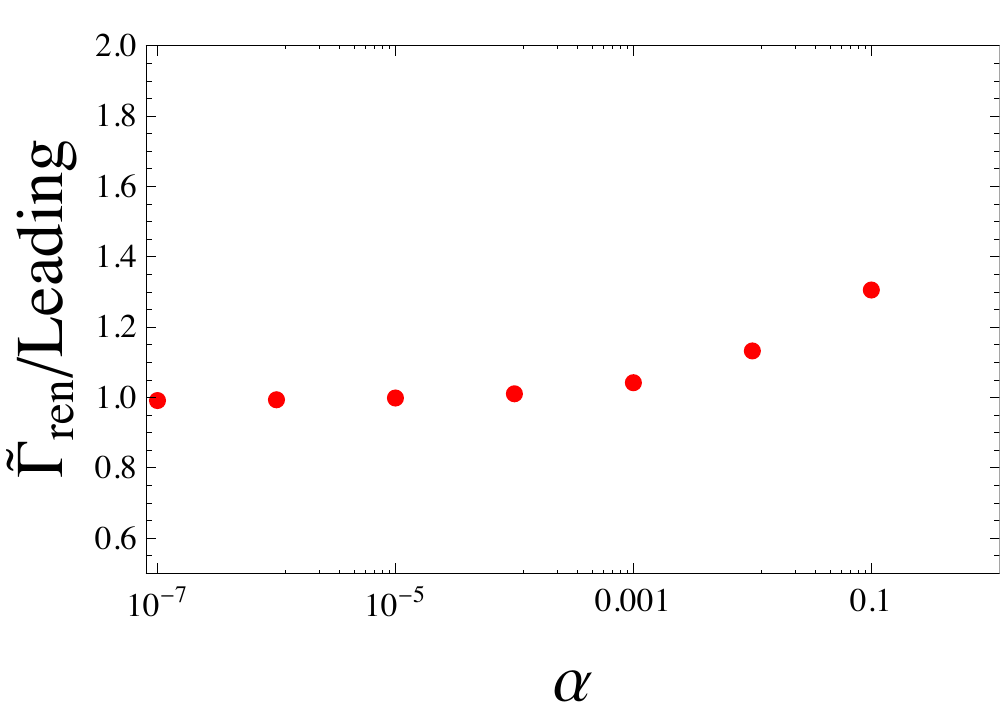}
\end{center}
\caption{The ratio of $\tilde\Gamma_{\rm ren}(m=0)$ vs. the leading asymptotic behavior of its two-point contribution
as a function of $\alpha$.}
\label{fig4}
\end{figure}

Figure~\ref{fig3} also suggests that $\tilde\Gamma_{\rm ren}(m)$ at $m=0$ diverges in the limit $\alpha \to 0$. Our perturbative results
of the previous section allow us to confirm and interpret this fact, and even to establish the precise asymptotic behavior of $\tilde\Gamma_{\rm ren}(m=0)$:
as we have seen, perturbatively all the $N$ - point contributions to the effective action are finite in the double limit $m,\alpha \to 0$ except for
the two-point function, and for the latter we have found the asymptotic small $\alpha$ behavior in (\ref{asympspin}). Thus we expect also the
full $\tilde\Gamma_{\rm ren}(m=0)$ to have the same small $\alpha$ behavior, 

\bear
\tilde\Gamma_{\rm ren}(m=0) \quad \stackrel{\alpha\to 0}{\sim} \quad
 \frac{1}{12}(\ln \alpha)^2 + \Bigl(-\frac{11}{72} + \frac{1}{2}\ln 2 - \frac{1}{6}\gamma_E \Bigr) \ln \alpha
\, .
\non
\label{Gammaasympspin}
\ear
This is confirmed by Fig.~\ref{fig4}, where we show a numerical plot of the ratio of the left and right hand sides of (\ref{Gammaasympspin}) as a function
of $\alpha$. 

\begin{figure}[t]
\begin{center}
\includegraphics[width=0.45\textwidth]{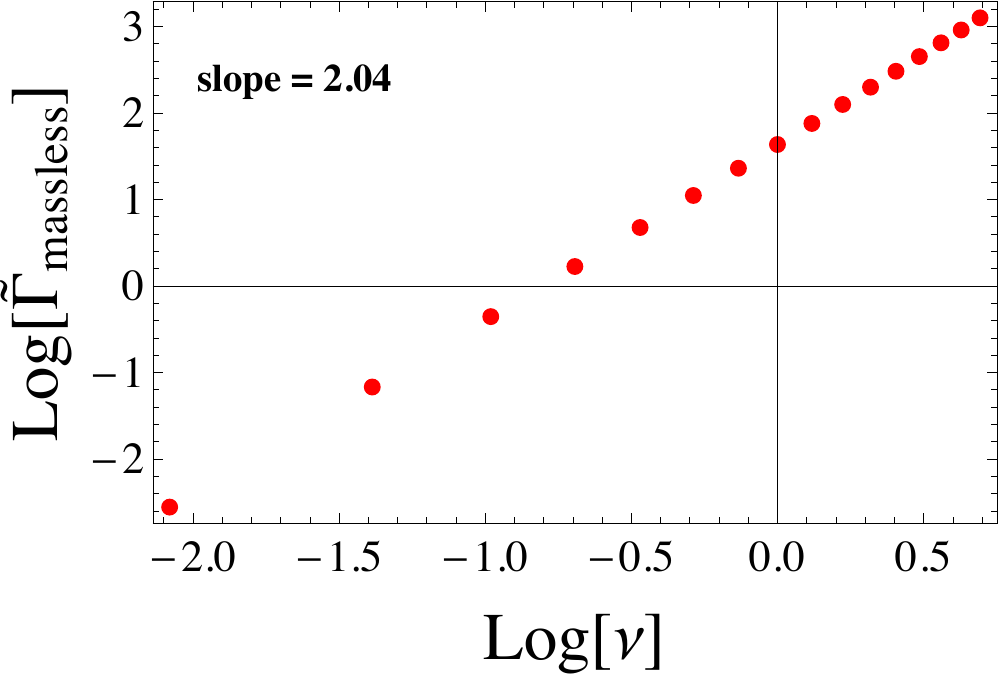}
\end{center}
\caption{The log of $\tilde\Gamma_{\rm ren}(m=0)$ as a function of $\log \nu$ for $\alpha =1/5000$.} 
\label{fig5}
\end{figure}

We can also trace the origin of the divergence of $\tilde\Gamma_{\rm ren}(0)$ for $\alpha\to 0$: it is easy to see that the 
$k$ - integral in our final result for the two-point function (\ref{gammafinal}) is, for $\alpha\to 0$, dominated by the region close to $0$. This implies that
this divergence is related to the divergence of the integral of the induced Maxwell term for $\alpha =0$ (see (\ref{alphazero})), of which
a finite part is still contained in the two-point function (for our choice of the unphysical renormalization condition $\mu =1$).
This fact that the small $\alpha$ divergence comes purely from the perturbative two-point contribution can be checked also in a
very different way: if the two-point contribution to the effective action becomes dominant over the higher-point ones for sufficiently small $\alpha$, then
also the dependence of the whole effective action on the background field normalization constant $\nu$ should become quadratic. In Fig. \ref{fig5}
we show that indeed numerically the $\nu$ - dependence of $\tilde\Gamma_{\rm ren}(m=0)$ for $\alpha = 1/5000$ becomes close to quadratic;
a fit for the exponent yields $2.04$.

\begin{figure}[t]
\begin{center}
\bigskip\bigskip
\includegraphics[width=0.45\textwidth]{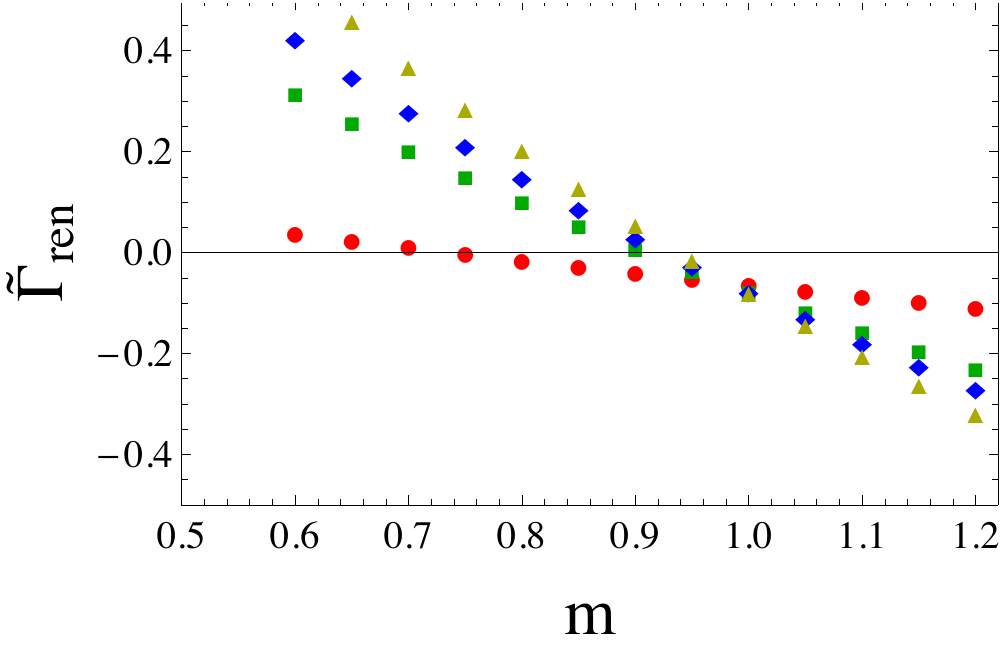}
\end{center}
\caption{The effective action $\tilde{\Gamma}_{{\rm ren}}$ for different values of $\alpha$. Dots correspond to $\alpha=1/10$, squares to $\alpha=1/100$, 
diamonds to $\alpha=1/200$ and triangles to $\alpha=1/450$.}
\label{fig6}

\end{figure}

One advantage of being able to calculate the effective action for the full range of masses is that one can look for zeros.
As seen from Fig.~\ref{fig2} for $\alpha=1/100$ $\tilde{\Gamma}_{{\rm ren}}(m)$ vanishes close to $m=1$. 
A more detailed study reveals that, remarkably, not only the existence but also the location of this zero seems to
be rather stable under variation of $\alpha$, as shown in Fig.~\ref{fig6}. 
These zeros of mass of $\tilde{\Gamma}_{{\rm ren}}(m)$ are shown 
in Table~\ref{tab1} for different values of $\alpha$.

\begin{table}
\begin{center}
\begin{tabular}{|c|c|}
\hline
$\alpha$ &  Crossing\\
\hline
 1/10 & 0.735540\\
1/100 & 0.907293 \\
1/200 & 0.925169\\
1/450 & 0.939393\\
\hline
\end{tabular}
\caption{Mass zero of the effective action as a function of $\alpha$.} 
\label{tab1}
\end{center}
\end{table}

\begin{figure}[t]
\begin{center}
\includegraphics[width=0.45\textwidth]{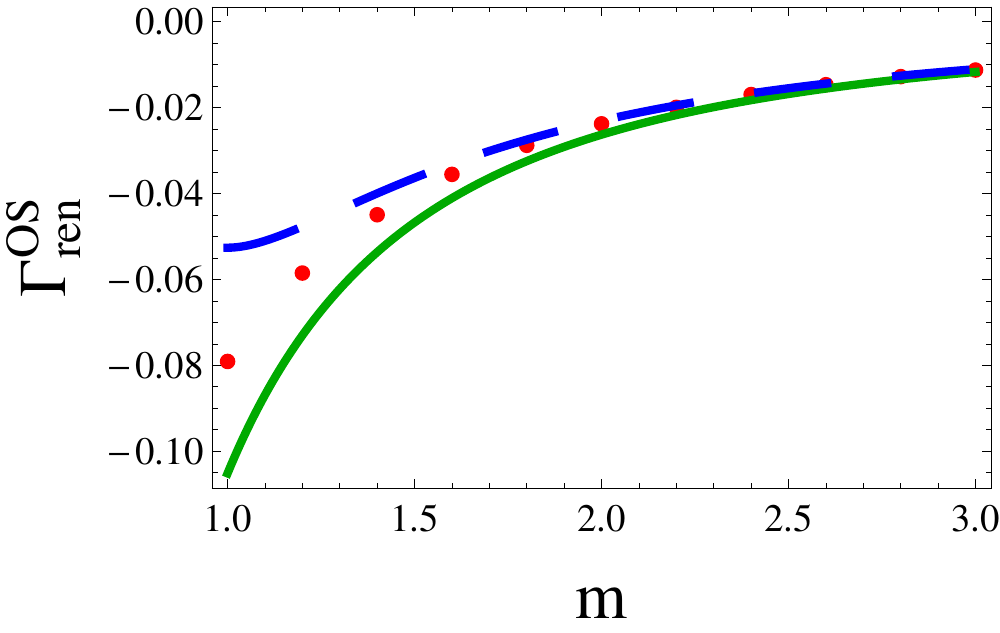}
\end{center}
\caption{The  effective action $\Gamma_{{\rm ren}}^{\rm OS}$ as a function of $m$ in the large-mass limit for $\alpha=1/100$. Dots represent the
exact effective action, 
the solid curve is produced by fitting terms only up to $1/m^{2}$ whereas the dashed curve corresponds to the behavior~(\ref{gammafit}).}
\label{fig7}

\end{figure}

 In the large-mass regime, we compare our numerical calculation of the physically renormalized
 effective action with its inverse mass (= heat kernel) expansion, using the leading and subleading terms in this
 expansion:

\begin{equation} \label{gammafit}
\Gamma^{\rm OS}_{\rm ren}(m) = \frac{c_{{\rm spin},2}}{m^2}+ \frac{c_{{\rm spin},4}}{m^4} + O\left(\frac{1}{m^6}\right) \, . 
\end{equation}
The coefficients $c_{{\rm spin},2}$  and $c_{{\rm spin},4}$ (which are still functions of $\alpha$) were given in (\ref{cspin24}).
As can be seen from Fig.~\ref{fig7},  the leading order approximation fits the numerical results very well in the large mass region,
and in an intermediate range of masses (between about $m=1.5$ and $m=2.0$) adding the subleading term leads to a better agreement with the
numerical data (in interpreting these results it should be kept in mind that, in applications of the inverse mass expansion,
typically any truncation to finite order breaks down completely at small enough masses, and adding a few terms more will
lower this point of breakdown only slightly; see, e.g., \cite{kwlemi} ). 

\section{Conclusions}
\label{conc}
\renewcommand{\theequation}{6.\arabic{equation}}
\setcounter{equation}{0}

The calculation of effective actions is a very important matter simply because the fermion determinant appears everywhere
in the standard model (see, e.g., \cite{hekosc,salcedo} and refs. therein). It is an exact contribution to the gauge field measure in the functional integral.
In this work, we have continued and extended the full mass range analysis of the scalar and spinor QED effective actions
for the $O(2)\times O(3)$ symmetric backgrounds (\ref{g2}), started in \cite{Dunne:2011fp}, by a more detailed 
numerical study of both the small and large mass behavior. 

In \cite{Dunne:2011fp}, only the unphysically renormalized versions $\tilde\Gamma_{\rm ren}(m)$ of these effective actions
were considered (corresponding to $\mu=1$), which are appropriate for the small mass limit, but have a logarithmic
divergence in $m$ in the large $m$ limit. This asymptotic logarithmic behavior was numerically well-reproduced in
\cite{Dunne:2011fp}, but its presence prevented one from probing into the physical part of the large mass expansion,
whose leading term is already $1/m^2$ - suppressed.
Here, we have instead used the physically renormalized effective actions $\Gamma_{\rm ren}^{\rm OS}(m)$ 
for the study of the large mass expansions. Going to  large 
masses demands  higher values of the cutoff $L$ and is computationally more challenging.  
However, we have  matched our numerical results not only against the leading term, 
but also against the subleading $O(1/m^4)$ term in the expansion. 
We have also calculated the expansion coefficients analytically for these backgrounds. 
 
At the intermediate mass range, we have demonstrated the ability of the method to compute zeroes of the effective action. 

Most of our effort  here has, however, gone into the study of the small-mass limit of the effective action. Our study of the
perturbative $N$ - point functions in this background has shown that, with the exception of the two-point function,
all of them are finite in the double limit $m,\alpha \to 0$ (the latter meaning the removal of the exponential IR suppression
factor). The two-point function is, for $\alpha >0$, made finite in the massless limit using the renormalization condition $\mu=1$.
Letting also $\alpha \to 0$ in it however produces an IR divergence whose $\alpha$ - dependence we have been able to calculate.
In our numerical study of the small mass limit of the effective action, we have improved on  \cite{Dunne:2011fp} by obtaining good numerical results
for $\tilde\Gamma_{\rm ren}(m)$ even at $m=0$, showing continuity for $m\to 0$  for various values of $\alpha$, and moreover
verifying that the full effective action at zero mass has the same diverging asymptotic behavior for $\alpha \to 0$ as its
two-point contribution. 

Our results further provide strong support for M. Fry's conjecture \cite{fry2006}
according to which the effective action for this type of background should, after the subtraction of its two-point and
four-point contributions, in the small-mass limit be dominated by a logarithmic divergence in the mass entirely due to the
chiral anomaly term. 
This term exists for the backgrounds (\ref{g2}) only at $\alpha=0$, which case is difficult to access with our method 
since, even after the subtraction of the true IR divergence contained in the two-point function,
one would still have spurious IR divergences in $\Gamma^{(\pm)}_{\rm ren}$ which will cancel
only in the sum of the low and high angular momentum contributions. 
This poses a formidable challenge for a numerical treatment.
Nevertheless, our results show that, as long as $\alpha >0$ and after the subtraction of the two-point function, 
the effective action is finite in the zero mass limit, both perturbatively and non-perturbatively. Given the finiteness of
the double limit $m,\alpha\to 0$ for all the $N$ - point functions but for the discarded two-point one, it is clear
that the appearance at $\alpha = 0$ of some term singular in the massless limit other than the known chiral anomaly one 
would signal some new nonperturbative effect different from, but similar to the chiral anomaly, 
which is hardly to be expected in QED at the one-loop level.

We believe that the work presented here not only provides an impressive demonstration of the power of the ``partial-wave-cutoff method'',
but also constitutes the most complete study performed so far of a one-loop QED effective action in a nontrivial background field. 

\medskip
\noindent
\textbf{Acknowledgements:}\\
We would like to thank M. Fry, G. Dunne and H. Min for discussions and comments on the manuscript.
N.A. and C.S. thank D. Kreimer and the Institutes of Physics and Mathematics, Humboldt-Universit\"at zu Berlin, for hospitality.
We acknowledge CIC-UMSNH and CONACyT grants. A. H also acknowledges support from Red-FAE CONACyT.

 \end{document}